\documentclass{article}
\usepackage[utf8]{inputenc}
\usepackage{amsmath}
\usepackage[colorinlistoftodos]{todonotes}
\usepackage{subcaption}

\newcommand{\tdab}[1]{} 

\graphicspath{{figs/}}

\title{A closed form for Jacobian reconstruction from timeseries and its application as an early warning signal in network dynamics}

\author{Edmund Barter$^{1*}$, Andreas Brechtel$^{2*}$, Barbara Drossel$^{2}$ \\[1.6mm] and Thilo Gross$^{3@}$}

\date{\small $^1$ \textit{University of Bristol, Department of Engineering Mathematics, Merchant Venturers Building, Woodland Road, Bristol BS8 1UB, UK}\\ $^2$ \textit{TU Darmstadt, Fachbereich Physik, Hochschulstrasse 12, 64289 Darmstadt, Germany}\\ $^3$ \textit{UC Davis, Department of Computer Science, 1 Shields Av. Davis Ca 95616, USA}\\
$@$ please address correspondence to thilo2gross@gmail.com\\ $*$ these authors contributed equally}

\begin{document}
\maketitle

\begin{abstract}
The Jacobian matrix of a dynamical system describes its response to perturbations.
Conversely one can estimate the Jacobian matrix by carefully monitoring how the system 
responds to environmental noise. Here we present a closed form analytical solution for the 
calculation of a system's Jacobian from a timeseries. Being able to access a system's Jacobian 
enables us to perform a broad range of mathematical analyses by which deeper insights into the system can 
be gained. Here we consider in particular the computation of the leading Jacobian eigenvalue 
as an early warning signal for critical transition. To illustrate this approach we apply it 
to ecological meta-foodweb models, which are strongly nonlinear dynamical multi-layer 
networks. Our analysis shows that accurate results can be obtained, although the data demand of the 
method is still high.  
\end{abstract}

\section{Introduction}
As humans we are dependent on the functioning of complex systems on many scales, ranging from our own body with its interlinked metabolic, signalling and microbial networks, via technical and organizational systems such as power grids and political systems, to the planetary-scale supply chains and the climate systems. All of these are complex nonlinear many variable systems, and as such at a risk of undergoing sudden, qualitative and potentially irreversible transitions \cite{scheffer2009critical}. Over the past decades there has been a steadily growing interest in methods that provide early warning of such transitions  \cite{scheffer2012anticipating,kuehn2015early}. 
In particular there is a growing awareness that such methods are needed for systems that are spatial or network based and hence 
inhherently high-dimensional \cite{kefi2014early,ludescher2014very,tirabassi2014interaction}.

The traditional approach to anticipating qualitative transitions in complex systems is mechanistic modelling. Many well-understood systems can be modelled so precisely that the model can accurately predict the threshold parameter values where transitions occur. For technical applications such as aircraft flight or stability of structures the model-based stability analysis is well established and in many cases part of a legally mandated licensing process \cite{raymer2018aircraft}. However, model-based approaches tend to produce poor results in systems that are less well known as seemingly minor details of the model can sometimes drastically affect transition points. 

It has long been known that critical transitions are generally preceded by critical slowing down \cite{stanley1987introduction,kuehn2011mathematical}. This phenomenon leads to a distinctive increase in the auto-correlation and cross-correlations of timeseries before a transition. The advantage of correlation-based warning signs is that detailed understanding of the system is not needed. Its disadvantage is that high frequency time series data are necessary for robust warning, which imposes a strong, and for many applications prohibitive, constraint.      

The model-based and correlation-based approaches to predicting critical transitions can be seen as two extreme strategies. The former is entirely based on structural knowledge of the system and uses real world data at most to fit model parameters. The latter eliminates the need for structural information, at the cost of a high time series data demand \cite{dakos2012methods,singh2018skill}. 

In many potential applications there are aspects of the system that are well understood because different variables are related via physical laws or subject to logical constraints \cite{singh2018skill}. If such bits of `structural' knowledge are available in a system it is desirable to exploit them for the construction of early warning signals. However, the same systems may also contain aspects that are considerably less well understood and hence make purely model based predictions unreliable. This defines the need for a middle way, where available structural information on a system is used to reduce the data demand, while time series data is used to close gaps in the understanding in areas where structural information is not available \cite{scheffer2012anticipating}.       

A common approach to finding a middle way in the construction of early warning signs is to use data assimilation approaches to continuously improve a dynamical model \cite{singh2018skill}. A classical application of these approaches is the eutrophication of shallow lakes \cite{carpenter1999management,boettiger2012early}, for which good early warning signs can be constructed using techniques such as Bayesian learning and Kalman filters \cite{peterson2003uncertainty,moradkhani2005dual,liu2001combined}. However, a broad investigation of such approaches shows that they may result in warning signs that are ``faint and late''\cite{singh2018skill}, while Boettiger and Hastings point out advantages of a model-based approach \cite{boettiger2012early}.

A different approach to critical transitions builds on linear stability analysis \cite{lade2012early,piovani2016linear}. In \cite{lade2012early}, the authors formulated a model that was sufficiently general to encompass the whole class of conceivable models into which a given system could potentially fall. Using the so-called generalized modelling approach, the Jacobian matrices were computed that govern the system's response to perturbations and from which the bifurcation points can be computed. Because the underlying models contain unknown functional relationships the Jacobian matrices that are thus obtained still contain unknown parameters. The authors used time series data to eliminate these remaining uncertainties. They showed that in this way accurate early warning signals can be constructed that only require very limited timeseries data.  

The present paper takes the idea of reconstructing a systems Jacobian from data one step further by avoiding assumptions on the functional form of the interaction terms. In a system that is subject to some noise, the cross correlations in timeseries encode very similar information to the Jacobian matrix of its deterministic backbone. Considered in isolation the cross correlations are not sufficient to reconstruct the full Jacobian, however a complete reconstruction is possible if we have some additional structural information. In systems that can be described as complex networks the network structure imposes constraints on which variables can interact directly, which in turn implies that some entries of the Jacobian must be zero. In a sufficiently sparse network, and particularly in multi-layer networks, knowing these zeroes provides sufficient information to reconstruct the remainder of the Jacobian from time series correlations.  

For illustration we apply Jacobian reconstruction approach to an ecological meta-foodweb model, formulated as a dynamical multi-layer network. 
By comparing with the known ground truth of the model, i.e., its exact Jacobian, we show that the Jacobian can be reconstructed faithfully and demonstrate its value as an early warning signal. We find that despite leveraging the structural information, the amount of timeseries data required for accurate results is at present still prohibitively high for the ecological application. However we discuss several avenues of future research that may reduce the data requirements to a point where the method becomes widely applicable.  

\section{Mathematical Background}
In every dynamical system that is in the vicinity of some form of long-term behaviour,
the response of the system to small perturbations in the variables can be captured by some matrix. In the simplest case the system is a system of ordinary differential equations (ODEs) 
\begin{equation}
\dot{\boldsymbol{x}} = f(\boldsymbol{x},\boldsymbol{p})    
\end{equation}
in which a vector of variables $\boldsymbol{x}$ evolves in time in a way that is dependent on a set of parameters $\boldsymbol{p}$. The simplest form of long-term behavior is rest in a stationary state $\boldsymbol{x}^*(\boldsymbol{p})$, such that
\begin{equation}
f(\boldsymbol{x}^*,\boldsymbol{p})=0
\end{equation}
The response to sufficiently small perturbations in the the steady state is then described by the Jacobian matrix $\bf J$, whose elements are computed as 
\begin{equation}
J_{ij} = \left. \frac{\partial}{\partial x_j} \dot{x}_i \right|_{\boldsymbol{x}=\boldsymbol{x}^*}     
\end{equation}
The steady state under consideration is stable when all eigenvalues of the Jacobian have negative real parts. A smooth change in the parameters $\boldsymbol{p}$ will generally cause the the eigenvalues to change smoothly. When such a change causes one or more eigenvalues to acquire positive real parts, then a bifurcation occurs in which the dynamics change qualitatively. 

Because the Jacobian is a real matrix its eigenvalues are real or form complex conjugate pairs. Hence there are two fundamental types of bifurcations. In bifurcations of fold-type a single real eigenvalue acquires a positive real part as it passes through zero. Several different forms of this bifurcation are commonly encountered in dynamical systems (fold bifurcation, pitchfork bifurcation, transcritical bifurcation), but generally speaking these are associated with a change in the number of steady states in the system or the exchange of stability properties. 

In a Hopf bifurcation a complex conjugate eigenvalue pair acquires a positive real part by crossing the imaginary axis of the complex plane. This bifurcation is generally associated with the onset of at least transient oscillations as the system departs from the steady state. 

Both types of bifurcations can occur in several different forms, some of which cause only relatively mild non-catastrophic transitions (say, replacing stationary behaviour with low amplitude oscillations), while others lead to a catastrophic (and potentially irreversible) departure from the steady state. 

One can distinguish between different types of bifurcations by computing normal form parameters that are functions of higher derivatives of $f$. However, this is beyond the scope of the present paper. 

In spatially extended systems, which are defined on a continuous space or on a spatial network of discrete nodes, the fundamental bifurcations can come in two flavors: A bifurcation may affect all points in space simultaneously in the same way, or it may affect points in space differently leading to the formation of spatial patterns. For clarity the bifurcations of the latter type are called Turing bifurcations (fold-like case) and wave instabilities (Hopf-like case).  

While not every bifurcation in a dynamical system is a critical transition, any bifurcation occurring in an important real world system is certainly a cause for concern. In this spirit our aim in the following is to reconstruct the Jacobian of a dynamical system from data in order to determine its leading eigenvalue. If the real part of this eigenvalue approaches zero, we interpret this as a warning signal for an impending bifurcation. 

\section{Jacobian Reconstruction}
Our aim in this section is to formulate a method that can reconstruct the Jacobian matrix of a dynamical system from time series. We do this by expanding on the work of Honerkamp \cite{honerkamp1990stochastische}, van Kampen \cite{vankampen2007stochastic}, and Steuer et al.~\cite{steuer2003observing}.  

We start from a stochastic timeseries that fluctuates around a steady state $\boldsymbol{x}^*$ of the underlying deterministic backbone of the system.   
As shown in \cite{honerkamp1990stochastische,vankampen2007stochastic,steuer2003observing} and reproduced in appendix \ref{appendixCovJ} the Jacobian $\bf J$ close to $\boldsymbol{x}^*$ is related to the covariance matrix $\bf \Gamma$ of timeseries, and to the fluctuation matrix $\bf D$ of the noise, via the equation  
\begin{align}
\label{eq:cvMatrix}
    \mathbf{J\Gamma}+\mathbf{\Gamma J}^T = -2\mathbf{D}.
\end{align}
In the following, we show that we can use this relationship to compute $\bf J$. Consider that a Jacobian matrix of linear dimension $N$ contains $N^2$ independent entries. By contrast, Eq.~(\ref{eq:cvMatrix}) equates two symmetric matrices, and hence imposes only $N(N+1)/2$ constraints on the elements of $\bf J$. Therefore in any application with multiple variables ($N>1$) the system is underdetermined such that we cannot recover the complete Jacobian purely from the time series.  

We can still recover the full Jacobian if we have additional information that we can leverage. Fortunately, in many applications some structural information is easily accessible. In particular in large spatially complex systems or reasonably sparse networks we know that certain variables cannot interact and hence the corresponding elements of the Jacobian must be zero. This yields a set of structural constraints of the form $J_{ij}=0$ for certain pairs $(i,j)$. If we can identify at least $N(N-1)/2$ such zero entries, then the timeseries contain enough information to reconstruct all remaining Jacobian entries. Below in Sec.~\ref{secDiscussion} we demonstrate that this is quite generally the case. 

Given a timeseries of the system's $N$ variables and a set of $G$ additional structural constraints, with $G\geq N(N-1)/2$ we now solve the Eq.~\ref{eq:cvMatrix} to estimate the nonzero entries of the Jacobian.  To understand how this equation is solved let us first consider the two-dimensional example
\begin{displaymath}
\small
    \left( \begin{array}{c c} J_{11} & J_{12} \\ J_{21} & J_{22} \end{array}  \right)
    \left( \begin{array}{c c} \Gamma_{11} & \Gamma_{12} \\ \Gamma_{12} & \Gamma_{22} \end{array}  \right)
    + 
    \left( \begin{array}{c c} \Gamma_{11} & \Gamma_{12} \\ \Gamma_{12} & \Gamma_{22} \end{array}  \right)
    \left( \begin{array}{c c} J_{11} & J_{21} \\ J_{12} & J_{22} \end{array}  \right)
    =
    -2 \left( \begin{array}{c c} D_{11} & D_{12} \\ D_{12} & D_{22} \end{array}  \right)
\end{displaymath}
which implies the independent conditions
\begin{eqnarray}
2J_{11}\Gamma_{11}+ 2J_{12}\Gamma_{12} &=& -2D_{11} \\  
J_{11}\Gamma_{12}+J_{12}\Gamma_{22}+J_{21}\Gamma_{11}+J_{22}\Gamma_{12} &=& -2 D_{12} \\
2J_{21}\Gamma_{12}+2J_{22}\Gamma_{22} &=& - 2D_{22}
\end{eqnarray}
with the second condition applying to both off-diagonal terms.
The left-hand side of these equations is a linear system. Hence
we can write the conditions in the form 
\begin{equation}
\label{eqBeq}
    {\bf B}\boldsymbol{j}=-2\boldsymbol{d}
\end{equation}
where $\bf B$ is a matrix, $\boldsymbol{j}$ is a column vector that contains the entries of the Jacobian, i.e.~$\boldsymbol{j}=(J_{11},J_{21},J_{12},J_{22})^T$ and $\boldsymbol{d}$ is the corresponding vector for $\bf D$. For the two dimensional example this reads
\begin{equation}
\left(\begin{array}{c c c c}
2 \Gamma_{11} & 0 & 2 \Gamma_{12} & 0 \\
\Gamma_{12} & \Gamma_{11} & \Gamma_{22} & \Gamma_{12} \\
\Gamma_{12} & \Gamma_{11} & \Gamma_{22} & \Gamma_{12} \\
0 &  2\Gamma_{12} & 0 & 2\Gamma_{22}  \\
\end{array}\right)
\left( \begin{array}{c }
J_{11}\\ J_{21}\\ J_{12} \\ J_{22}
\end{array}\right) 
=-2 
\left(\begin{array}{c}
D_{11}\\ D_{12}\\ D_{12} \\ D_{22}
\end{array}
\right)   
\end{equation}
The form of this equation suggests that we can solve it for $\boldsymbol j$ by multiplying ${\bf B}^{-1}$ from the left. However, we have to take care because $\bf B$ is not invertible because the two center rows are identical, which is a consequence of the missing information. We can fix this problem by using the additional constraints. Imposing structural constraints on the two-variable system makes this example almost pointless, but, for the purpose of illustration, let us assume that we know that variable 1 cannot depend on variable 2 and hence $J_{12}=0$. We can represent this constraint in the same matrix equation as the system by adding it as an additional line,
\begin{equation}
\label{eqMatrixForm}
\left(\begin{array}{c c c c}
2 \Gamma_{11} & 0 & 2 \Gamma_{12} & 0 \\
\Gamma_{12} & \Gamma_{11} & \Gamma_{22} & \Gamma_{12} \\
\Gamma_{12} & \Gamma_{11} & \Gamma_{22} & \Gamma_{12} \\
0 &  2\Gamma_{12} & 0 & 2\Gamma_{22}  \\
0 & 0 & 1 & 0 
\end{array}\right)
\left( \begin{array}{c }
J_{11}\\ J_{21}\\ J_{12} \\ J_{22}
\end{array}\right) 
=-2 
\left(\begin{array}{c}
D_{11}\\ D_{12} \\ D_{12} \\ D_{22} \\ 0
\end{array}
\right)   
\end{equation}
If this is the only constraint then we can now drop the third row of the matrix and the third entry of the vector on the right-hand side and solve for $\boldsymbol{j}$ by matrix inversion. In practice there are typically additional constraints and hence the system will be overdetermined. In this case we use least squares optimization to find an approximate solution. As we will see below the form of Eq.~(\ref{eqMatrixForm}) is very convenient for finding the least squares solution. In particular it allows us to obtain a analytical closed form solution for $\boldsymbol{j}$.

Let us now generalize from the two-dimensional example to systems with an arbitrary number of variables. For this purpose we define the vectorization operator \cite{magnus2019matrix} 
\begin{equation}
    {\rm vec}({\bf X}) = (X_{11},X_{21},\ldots X_{N1},X_{12,}\ldots )^{\rm T}
\end{equation}
We can now write Eq.~(\ref{eqBeq}) as
\begin{equation}
    {\bf B} \, {\rm vec} ({\bf J}) = -2 \, {\rm vec} ({\bf D})
\end{equation}
i.e.~the vectorization of a matrix is the columns of the matrix stacked on top of each other.
To find the general form of ${\bf B}$ we start from Eq.~(\ref{eq:cvMatrix}) and vectorize both sides, which yields 
\begin{equation}
    {\rm vec}(\mathbf{J\Gamma}+\mathbf{\Gamma J}^{\rm T}) = {\rm vec}(-2{\bf D}).
\end{equation}
Because vectorization is a linear operator we can pull the -2 out of the vectorization on the right hand side and apply the vectorization separately to the two terms on the left hand side, hence 
\begin{equation}
\label{eqHalfWayThere}
    {\rm vec}(\mathbf{J\Gamma})+{\rm vec}(\mathbf{\Gamma J}^{\rm T}) = -2 \, {\rm vec}({\bf D}).
\end{equation}
It is known \cite{magnus2019matrix} that for matrices ${\bf X},{\bf Y},{\bf Z}$ the following identity holds (cf.~appendix \ref{appendixVecKronecker}):
\begin{equation}
\label{eqKroneckerLink}
{\rm vec}({\bf X}{\bf Y}{\bf Z}) = ({\bf Z}^{\rm T}\otimes {\bf X}) \, {\rm vec}({\bf Y}) 
\end{equation}
where $\otimes$ is the Kronecker product of matrices defined by
\begin{equation}
{\bf X} \otimes {\bf Y} = \left(\begin{array}{c c c} 
X_{11} {\bf Y} & X_{12} {\bf Y} & \ldots \\
X_{21} {\bf Y} & X_{22} {\bf Y} & \ldots \\
\vdots & \vdots & \ddots
\end{array} \right)    
\end{equation}   
We now substitute ${\bf Y}={\bf J}$, ${\bf Z}={\bf \Gamma}$ and ${\bf X}={\bf I}$, where ${\bf I}$ is the identity matrix of appropriate size. This yields 
\begin{equation}
\label{eqfirstterm}
{\rm vec}({\bf J}{\bf \Gamma})={\rm vec}({\bf I}{\bf J}{\bf \Gamma}) = ({\bf \Gamma}\otimes {\bf I}) \, {\rm vec}({\bf J})    
\end{equation}
which brings the first term from Eq.~(\ref{eqHalfWayThere}) into the desired form. If we try the same for the second term we find
\begin{equation}
\label{eqSecondTerm}
{\rm vec}({\bf \Gamma}{\bf J}^{\rm T})={\rm vec}({\bf \Gamma}{\bf J}^{\rm T}{\bf I}) = ({\bf I}\otimes{\bf \Gamma} ) \, {\rm vec}({\bf J^{\rm T}})    
\end{equation}
which is not quite the desired form because vectorization of the transpose of $\bf J$ appears rather than the vectorization of $\bf J$ itself. The vectorization of a matrix and the vectorization of its transpose are not identical but closely related. Consider that for our two-dimensional the two vectorizations are related by
\begin{equation}
{\rm vec}({\bf J}^{\rm T}) = \left(\begin{array}{c}
J_{11}\\ J_{12}\\ J_{21} \\ J_{22} 
\end{array} \right) = {\bf C} \left(\begin{array}{c}
J_{11}\\ J_{21}\\ J_{12} \\ J_{22} 
\end{array} \right) 
= {\bf C} \, {\rm vec}({\bf J})
\end{equation}
where 
\begin{equation}
\label{eqPermutationI}
    {\bf C}=\left(\begin{array}{c c c c}
    1 & 0 & 0 & 0 \\
    0 & 0 & 1 & 0 \\
    0 & 1 & 0 & 0 \\
    0 & 0 & 0 & 1
    \end{array} \right)
\end{equation}
is a permutation matrix. In the general case we can still write 
\begin{equation}
\label{eqPermutationII}
  {\rm vec}({\bf J}^{\rm T}) = {\bf C} \, {\rm vec}({\bf J})  
\end{equation}
where $\bf C$ is now a permutation matrix of size $N^2 \times N^2$, one can construct this matrix from blocks of size $N\times N$ as
\begin{equation}
{\bf C} = \left(\begin{array}{c c c}
{\bf C_{11}} & {\bf C_{12}} & \ldots \\
{\bf C_{21}} & {\bf C_{22}} & \ldots \\
\vdots & \vdots & \ddots 
\end{array} \right)
\end{equation}
Here, $\bf C_{nm}$ is a $N\times N$-matrix defined by
\begin{equation}
   ({\bf C_{nm}})_{ij} = \delta_{im}\delta_{nj} 
\end{equation}
where $\delta$ is the Kronecker delta. 

Using the commutation matrix we can write Eq.~(\ref{eqSecondTerm}) as 
\begin{equation}
{\rm vec}({\bf \Gamma}{\bf J}^{\rm T})= ({\bf I}\otimes{\bf \Gamma} ) \, {\rm vec}({\bf J^{\rm T}})  = ({\bf I}\otimes{\bf \Gamma} ) \, {\bf C} \, {\rm vec}({\bf J})
\end{equation}
Substituting this relationship and Eq.~(\ref{eqfirstterm}) into 
Eq.~(\ref{eqHalfWayThere}) we get the desired form
\begin{equation}
\label{eqAlmostThere}
    \underbrace{\left( ({\bf \Gamma}\otimes {\bf I}) + ({\bf I}\otimes{\bf \Gamma} ) \, {\bf C} \right) }_{\bf B} \underbrace{{\rm vec}({\bf J})}_{\boldsymbol{j}}  = -2 \underbrace{{\rm vec}({\bf D})}_{\boldsymbol{d}}.
\end{equation}
We now know how to construct the matrix $\bf B$ such that we can write the system in the form 
\begin{equation}
\label{eqBthesecond}
{\bf B}\boldsymbol{j} = -2 \boldsymbol{d}\, .
\end{equation}
However, this system is still underdetermined. If we know that some elements of $\boldsymbol{j}$ must be zero we can represent this knowledge in a matrix $\bf U$. For this purpose we can gather the respective elements of $\boldsymbol{j}$ in an ordered list $\mathcal{U}$ and then define $\bf U$ as an $|\mathcal{U}|\times N$ matrix with
\begin{equation}
U_{nm} = \left\{\begin{array}{l l} 
1 &\quad \mbox{if $j_m$ is the $n$'th entry of $\mathcal{U}$ } \\
0 &\quad \mbox{otherwise}
\end{array}\right.    \label{def:U}
\end{equation}
Each row of this matrix represents one of the constraints that we wish to impose. For example if in the fourth row the only nonzero entry is in the 8th column this means our forth condition is that the 8th element of $\boldsymbol{j}$ must be zero. 

In analogy to the small example we can now impose the additional conditions on the system by stacking them below the matrix $\bf B$ such that Eq.~(\ref{eqBthesecond}) becomes
\begin{equation}
\left(\begin{array}{c}{\bf B}\\ {\bf U}\end{array}\right)\boldsymbol{j} = -2 \left(\begin{array}{c} \boldsymbol{d} \\ \boldsymbol{0}\end{array} \right)
\end{equation}
where $\boldsymbol 0$ is a column vector containing $|\mathcal{U}|$ zeroes. The () that appear in this equation should be read as a block-wise notation for matrices/vectors, where rows are stacked on top of each other in analogy to Eq.~(\ref{eqMatrixForm}). 

To simplify the notation we introduce
\begin{equation}
\hat{\bf B } =  \left(\begin{array}{c}{\bf B}\\ {\bf U}\end{array}\right)   
\quad
\hat{\bf \boldsymbol{d} } =  \left( \begin{array}{c}\boldsymbol{d} \\ \boldsymbol{0}\end{array} \right)
\end{equation}
which allows us to write the whole set of conditions once again in the form
\begin{equation}
\hat{\bf B }\boldsymbol{j} = \hat{\boldsymbol{d}}
\end{equation}
In practise this will now be an overdetermined system, such that no exact solution exists. However, finding an approximation that minimizes the squares of the deviations in each row is a well-known problem. The known solution \cite{strang1976linear,trefethen1997numerical} (cf.~appendix \ref{appendixPseudoinverse}) for this problem  is
\begin{equation}
\label{eqTheAnswer}
\boldsymbol{j} = ({\hat{\bf B}}^{\rm T}\hat{\bf B})^{-1}{\hat{\bf B}}^{\rm T}\hat{\boldsymbol{d}}     
\end{equation}
where the expression $({\hat{\bf B}}^{\rm T}\hat{\bf B})^{-1}{\hat{\bf B}}^{\rm T}$, the pseudoinverse of $\hat{\bf B}$,appears. 

The equation provides a closed form solution by which the Jacobian elements can be computed from the covariance matrix of the timeseries, a known or estimated fluctuation matrix, and a set of additional structural constraints on the Jacobian.  
 
Typically the least squares fit will not set the Jacobian elements governed by the structural constraints exactly to zero. One may therefore enforce these known zeros by setting the respective elements of the Jacobian explicitly to zero after the computation of Eq.~(\ref{eqTheAnswer}) has finished. In numerical experiments, described below, we found that this improved the accuracy of Jacobian eigenvalues estimated by this method. 
 
To summarize, the Jacobian $\bf J$ of an $N$-dimensional system close to a steady state can be reconstructed as follows:
\begin{enumerate}
    \item Compute the co-variance matrix from the time series data, $\Gamma_{ij}=\left<X_iX_j\right>$.
    \item Construct the diagonal fluctuations matrix $\mathbf{D}$, and compute $\boldsymbol{d}={\rm vec}({\bf D})$. In some systems these fluctuations can be measured directly, otherwise a reasonable approximation may be derived based on assumptions on the underlying noise process \cite{vankampen2007stochastic}. 
    \item Construct the permutation matrix $\bf C$ according to Eqs.~(\ref{eqPermutationI},\ref{eqPermutationII}).
    \item Compute the matrix 
    \begin{displaymath}{\bf B}=\left({\bf \Gamma} \otimes \mathbf{I} +  \left( \mathbf{I} \otimes {\bf \Gamma} \right) \mathbf{C} \right),\end{displaymath}
    where $\bf I$ is the $N\times N$ identity matrix.  
    \item Define $\boldsymbol{j}={\rm vec}(\bf J)$ and identify at least $N(N-1)/2$ elements of $\boldsymbol{j}$ that must be zero due to structural constraints. Use these to construct the matrix $\bf U$ (see Eq.~\eqref{def:U}). The column-dimension of $\bf U$ is the row dimension of $\boldsymbol{j}$, and 
    row dimension of $\bf U$ is identical to the number of structural constraints. The matrix has exactly one non-zero entry in each row such that $U_{nm}=1$ if the $n$th structural condition reads $j_m=0$.
    \item Construct 
    \begin{displaymath}
     \hat{\bf B } =  \left(\begin{array}{c}{\bf B}\\ {\bf U}\end{array}\right) \quad  
    \hat{\bf \boldsymbol{d} } =  \left( \begin{array}{c}\boldsymbol{d} \\ \boldsymbol{0}\end{array} \right)
    \end{displaymath}
    \item Compute 
    \begin{displaymath}
    \boldsymbol{j} = ({\hat{\bf B}}^{\rm T}\hat{\bf B})^{-1}{\hat{\bf B}}^{\rm T}\hat{\boldsymbol{d}}    
    \end{displaymath}
    and recover the Jacobian $\bf J$ from its vectorization $\boldsymbol{j}$.
    \item Set the elements of $\bf J$ governed by structural constraints explicitly to zero. 
\end{enumerate}

\section{Application to a meta-foodweb model}
In the following we explore if the Jacobian matrix can be reconstructed sufficiently accurately to warn of impending critical transitions. For this purpose, we constructed a test system of realistic complexity for which the Jacobian matrix is nevertheless known analytically, and we created noise timeseries.   

We used a meta-foodweb model already studied in \cite{brechtel2018master,brechtel2019farranging} (see appendix \ref{appendixFWmodel} for details).
The model consists of a spatial network of $P$ habitat patches linked by avenues of species dispersal (Fig.~1). Each patch harbours a complex foodweb, consisting of $S$ nodes that represent populations of different species, which are linked by predator-prey interactions. The dynamics of the system are given by a set of differential equations that govern the changes in variables due to diffusive dispersal between patches and biological processes occurring within a patch (primary production, predator-prey interaction, natural mortality).  
 
 \begin{figure}[t]
\includegraphics[width=\textwidth]{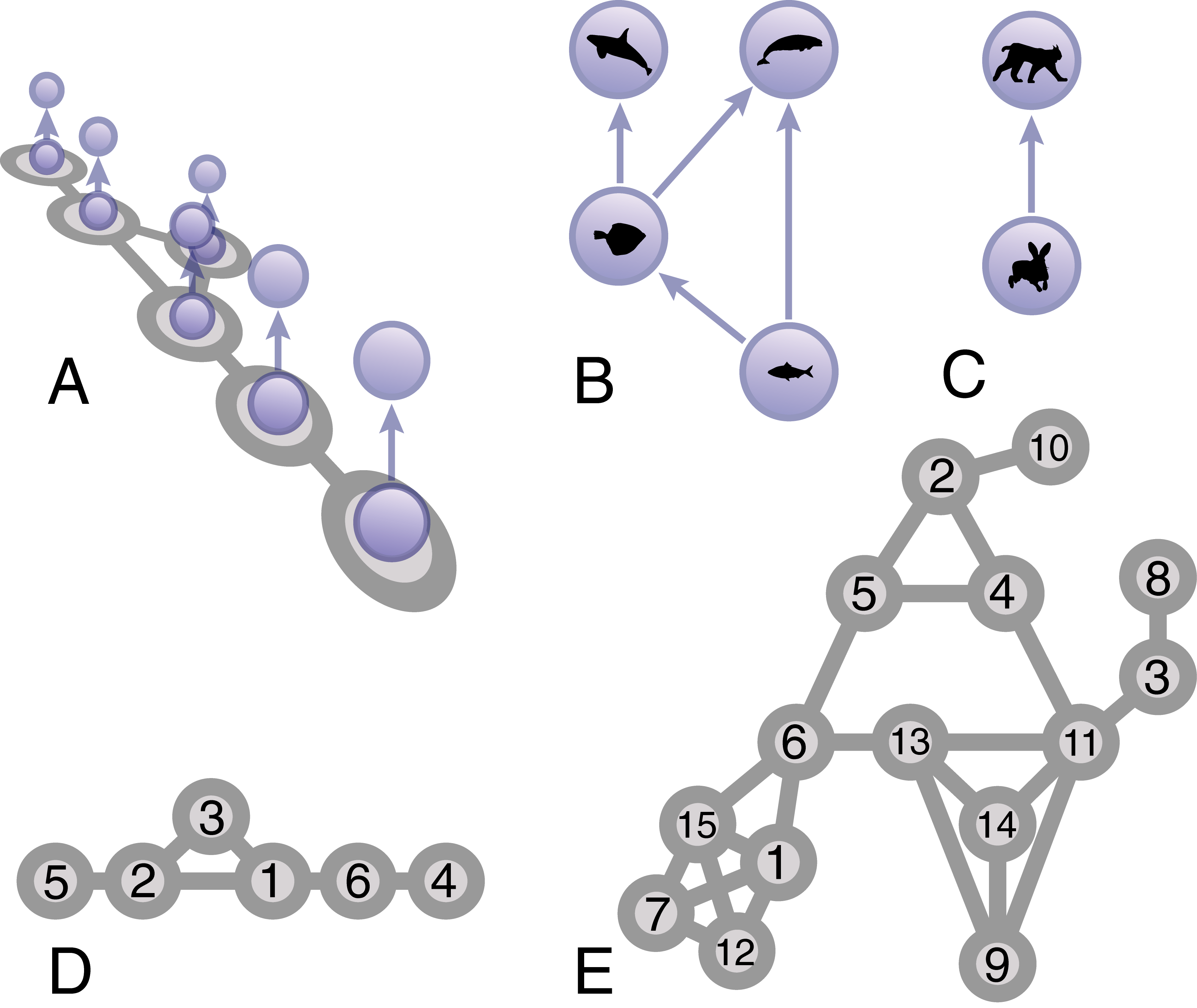}
\caption{Schematic representation of the model system. We consider a multilayer network where copies of an ecological food web exist in different geographical patches (A). We consider an intra-guild predation food web with an additional predator (B) and a predator-prey system (C). For the spacial network, we use 
the smallest completely asymmetric graph (D) and a larger random topology, generated as a random geometric graph(E).}
\end{figure}
 
The model contains several complicating factors encountered in real systems. The functions used to model the biological processes are strongly nonlinear.
They capture various saturation effects and realistic responses to the availability of different food sources (prey switching).
The dynamics of different species occurs on different time scales according to biological scaling relationships which relate a species position in the foodweb to its expected biomass turnover rate \cite{brose2006allometric}. Similar scaling relationships also govern the rate at which different species disperse across the spatial network\cite{brechtel2018master}.  
 
We consider two different versions of the patch topology. The smaller of the two consists 6 patches which are connected in such a way that they form the smallest completely asymmetric network (Fig.~1D). The larger one contains 15 patches and was generated as a random geometric graph (Fig.~1E). It has a comparatively large diameter and high clustering and contains several symmetries that are characteristic for this type of spatial networks \cite{nyberg2015mesoscopic}.
For the food webs we used a predator-prey system  consisting of two species (Fig.~1C), as well as a four-species system in the shape of the so-called intra-guild predation motif (Fig.~1B), a common and well-studied foodweb motif.  
 
Work by Nakao and Mikhailov \cite{nakao2010turing} and the extension of their approach to meta-foodwebs \cite{brechtel2018master} showed that network models behave analogously to dynamical systems in continuous space. Hence the theory of pattern formation in partial-differential equation systems can transferred almost exactly to these dynamical networks. This means that our test systems can exhibit instabilities that are best described as pattern-forming bifurcations, specifically Turing and Wave instabilities. Pattern forming instabilities in predator-prey systems in continuous space were studied in detail in \cite{baurmann2007instabilities} which made it easy to locate these bifurcations also in our multi-layer network predator-prey system (Fig.~2).  

\begin{figure}[ht]
\begin{center}
\includegraphics[scale=1]{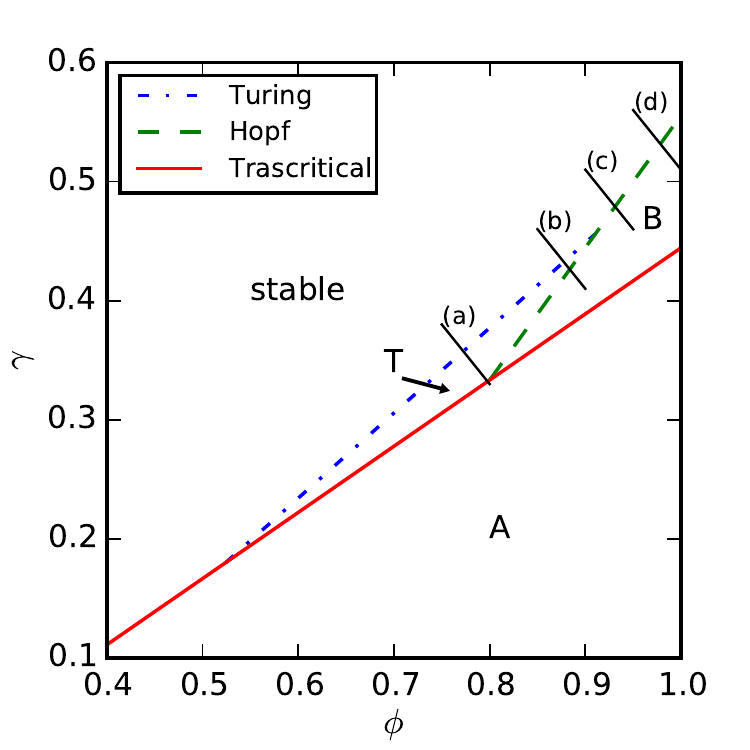}
\caption{Bifurcation diagram of the predator-prey system on the six patches topology. Stability of the state under consideration changes in response to changes in the sensitivity of biomass production to producer biomass $\phi$ and the sensitivity of predation to prey biomass $\gamma$. The state under consideration is stable in the top left area. Stability is lost when either of three bifurcations occur (Turing, Transcritical, Hopf). After the loss of stability the system approaches a state of homogeneous oscillations (A), a different homogeneous stationary state (B), or a state of stationary patterns (T). The bifurcation diagram was computed using the master stability function approach from \cite{brechtel2018master} (see appendix). It corresponds directly to Fig.~1 from \cite{baurmann2007instabilities} which studies a predator-prey system in continuous space. Lines (a-d) indicate the transects used for the corresponding simulations in Fig.~3.}
\end{center}
\end{figure}
\begin{figure}[htb!]
\includegraphics[width=\textwidth]{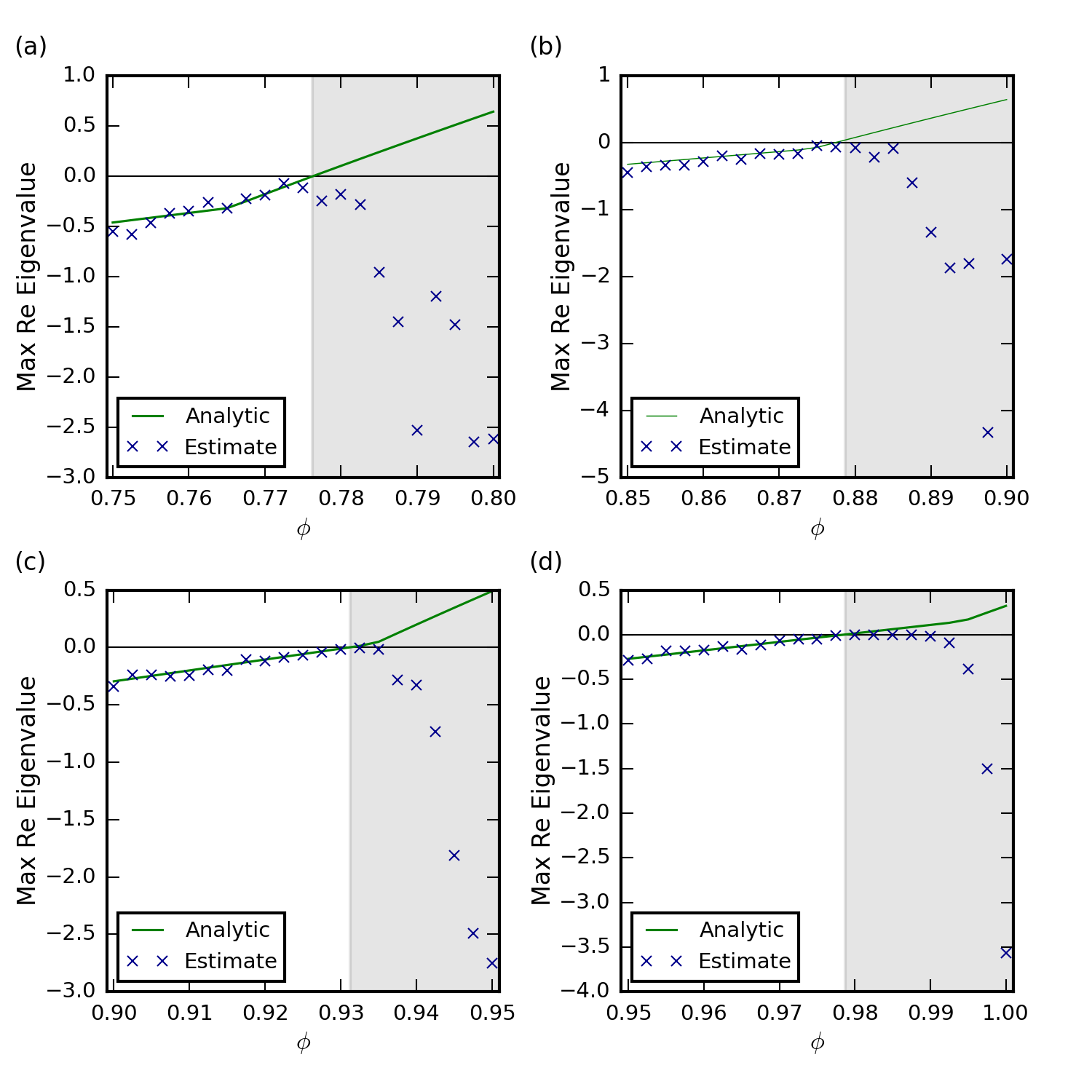}
\caption{Comparison of the analytical ground truth for the leading eigenvalue with an eigenvalue estimate from Jacobian reconstruction. The panels correspond to the 4 transects shown in Fig.~2. Estimates are in good agreement with the analytical value until stability is lost and the timeseries depat from the steady state. [Parameters $\phi,\gamma$: a) 0.75, 0.38 to 0.8, 0.33; b) 0.85, 0.46 to 0.9, 0.41; c) 0.9, 0.51 to 0.95, 0.46; d) 0.95, 0.56 to 1.0, 0.51. Bifurcations encountered are Turing (a,b) and Hopf (c,d) ]
}
\end{figure}

Using a generalized modelling approach \cite{gross2006generalized,gross2009generalized,plitzko2012complexitystability} we analytically computed the Jacobian that describes the Jacobian matrices of the two example food webs on arbitrary spatial topologies. We then picked specific realizations of models in which relevant bifurcations occurred. To produce noisy timeseries we simulated these models with added noise using the Euler-Maruyama method (see appendix \ref{appendixEMmethod}).

\section{Results}
As a first test we generated simulated noisy timeseries 
containing $2\cdot 10^5$ data points from the two-species system on the six-patch topology. Parameter values for these simulations were chosen to lie on four transects through the parameters space that crossed bifurcation points. For each of these time series we then reconstructed the Jacobian matrix and computed the leading eigenvalue. Before the bifurcation point the estimated eigenvalues are in very good agreement with the known ground truth provided by the analytic eigenvalues (Fig.~3). 

\begin{figure}[htb!]
\centering
\includegraphics[scale=1]{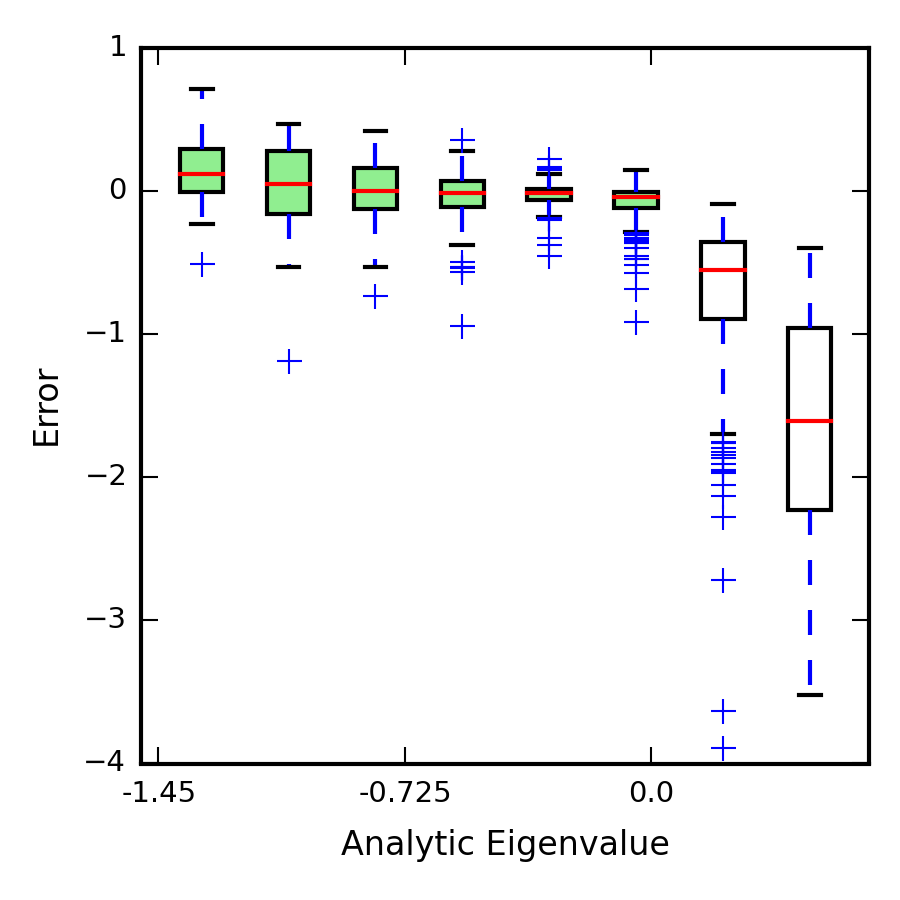}
\caption{The plot shows the error of the eigenvalue approximation in dependence to the analytic eigenvalue for the $S=2$ species on $N=6$ patch predator prey system. The error is the difference of the estimated eigenvalue and the analytic one. To gather the data 15 parameter trajectories covering transcritical, Hopf and Turing bifurcations where sampled 3 times each. The errors are small for negative leading eigenvalues. The errors for greater analytic eigenvalues where cut, because the estimation is not expected to work for positive eigenvalues.}
\end{figure}

As we follow the transects the real part of the leading eigenvalue crosses zero in a bifurcation. When the bifurcation occurs the reconstructed eigenvalue departs from the analytical value. This behavior is expected as the analytical solution continues to show the eigenvalues around the, now unstable, steady state, whereas the reconstruction algorithm computes the leading eigenvalue associated with the new dynamics, which has now departed from the previous steady state. 

We  note that in transect d) the reconstructed eigenvalue is almost exactly zero in a wide region after the bifurcation. This happens because the system approaches a stable limit cycle for which the leading Lyapunov exponent is zero. The recovery of this zero by the algorithm provides some (unexpected) evidence suggesting that the method reveals some salient information even in non-stationary states. 

Careful examination of the transects (Fig.~3) suggests that the accuracy of reconstruction improves as we approach the bifurcation point. To explore this further we generated 15 transects in the vicinity of bifurcation points and generated three sets of timeseries along every one of the transects (see appendix for details). The results confirm that the accuracy of the estimate improves as the bifurcation point is approached (Fig.~4).

\begin{figure}[htb]
\includegraphics[width=\textwidth]{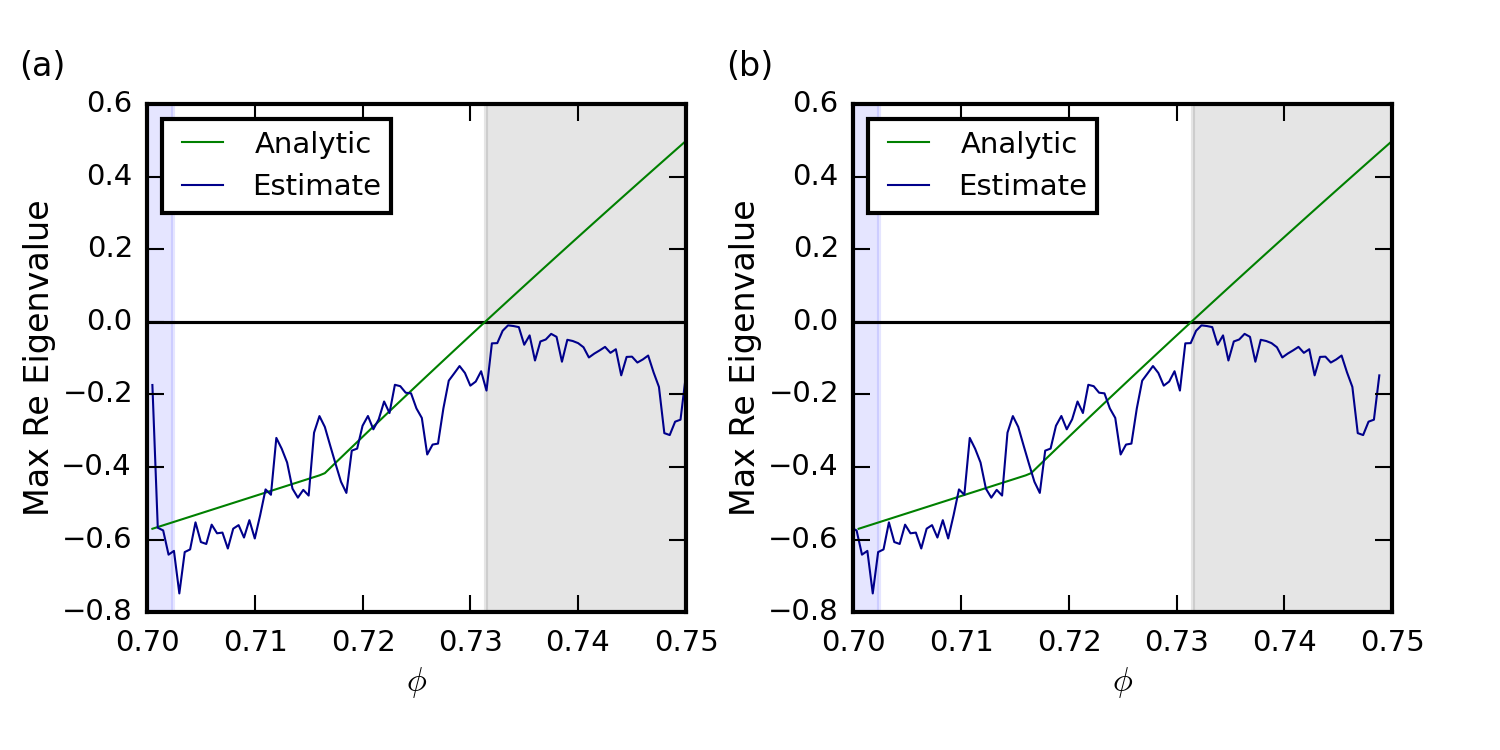}
\caption{Eigenvalue estimation in a system with continuously changing parameters. We compare the estimated eigenvalue from Jacobian reconstruction in a sliding window to the analytic ground truth for a predator-prey system on 6 geographical patches close to a Turing bifurcation. The panels show the same result, however the estimate is placed at the end of the observation window (panel a) or in the center (panel b). For large $\phi$ the steady state described by the analytical eigenvalues is unstable (greyed region) and hence the reconstruction yields eigenvalues of a different state. The width of the sliding window is indicated on the left (shaded blue).
[Parameters: $\phi,\gamma$ is changed from 0.70,0.35 to 0.75,0.30]}
\end{figure}

We now consider the case where parameters are slowly changing over a long simulation run. Our aim is to estimate the leading eigenvalue of the Jacobian over time as this slow change in the system is taking place. For this purpose we apply the proposed method to reconstruct the Jacobian in sliding time window of length $\tau$. The results (Fig.~5) show that the estimates based on the sliding window are more noisy, undergoing visible fluctuations around the true value. Nevertheless the trend of the eigenvalue approaching zero is still clearly captured.  

\begin{figure}[htb!]
    \includegraphics[width=\textwidth]{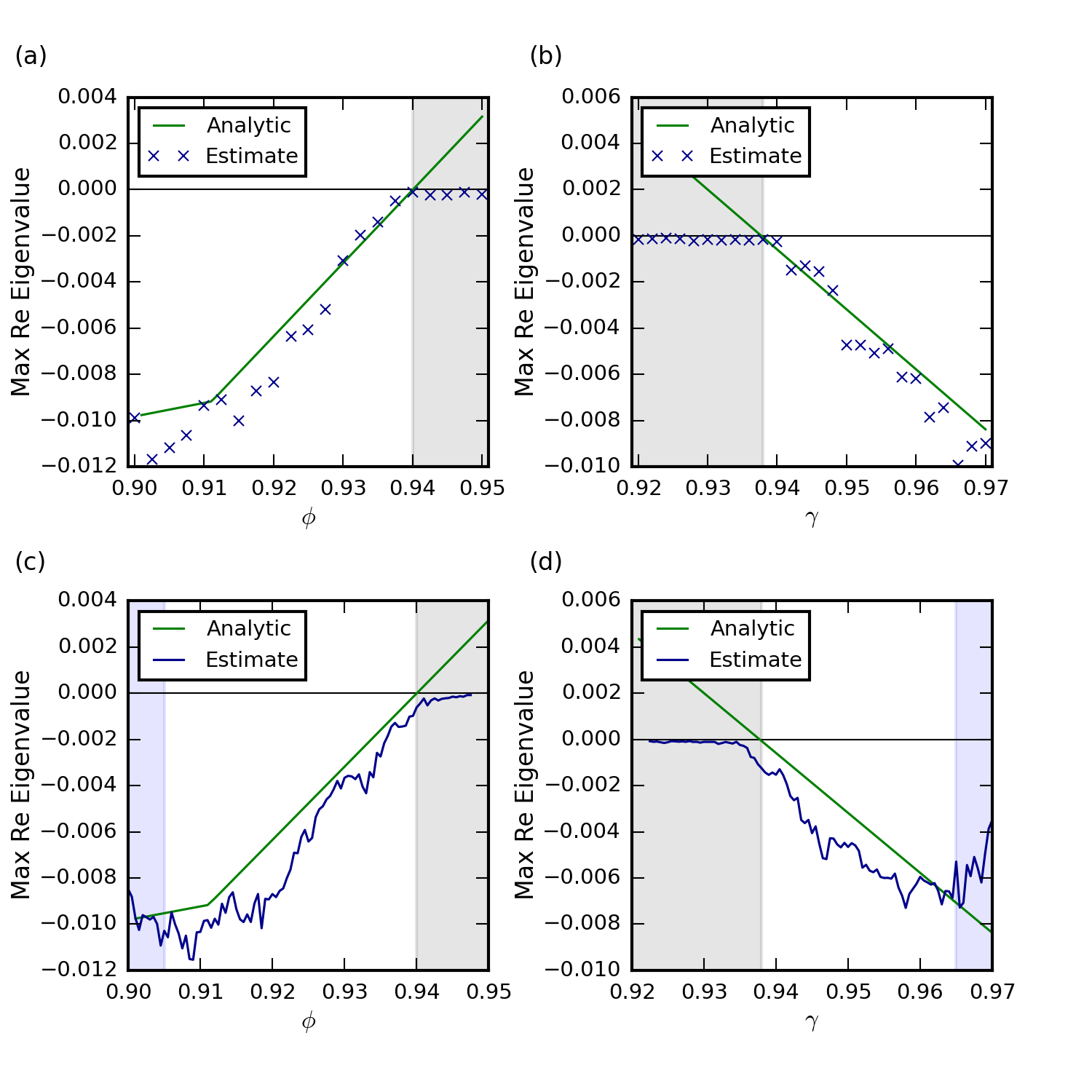}
    \caption{Eigenvalue estimations for the system with 4 species on 15 patches.
    We consider loss of stability as a result of the change of two different parameters, governing the nonlinearity of primary production, $\phi$ (panel a,c) and the nonlinearity of the functional response of attack rate to prey density, $\gamma$ (b,d). The leading eigenvalue of the Jacobian was reconstructed from timeseries for fixed parameters (a,b) and parameter transsects (c,d). The steady state under consideration is unstable subsequent to a Hopf bifurcation (region shaded grey). For the transects estimates are shown in the center of the sampling window (window width is indicated by blue area). See appendix for parameters and details.}
    \label{fig:bigger_system}
\end{figure}

To test the applicability to larger networks, we apply Jacobian reconstruction to a model system with 4 species on 15 patches. Even in this larger system the accuracy is still quite good but the estimated eigenvalue is systematically slightly less than the true value. For the case of fixed parameter values the difference once again disappears as we approach the bifurcation. But for the case of sliding parameter values a small difference remains. We suspect that this may be the combined effect of the non-autonomous nature of this systems and the noise leading to bifurcation delay. This delay effect can be expected to be more pronounced in the larger food web due to the presence of higher-level predator whose dynamics happen on correspondingly longer timescales. If this is the case then the reconstructed Jacobian eigenvalue may actually offer a better estimate of the relevant transition point than the analytical solution for the system without noise.  

\begin{figure}[htb!]
    \includegraphics[width=\textwidth]{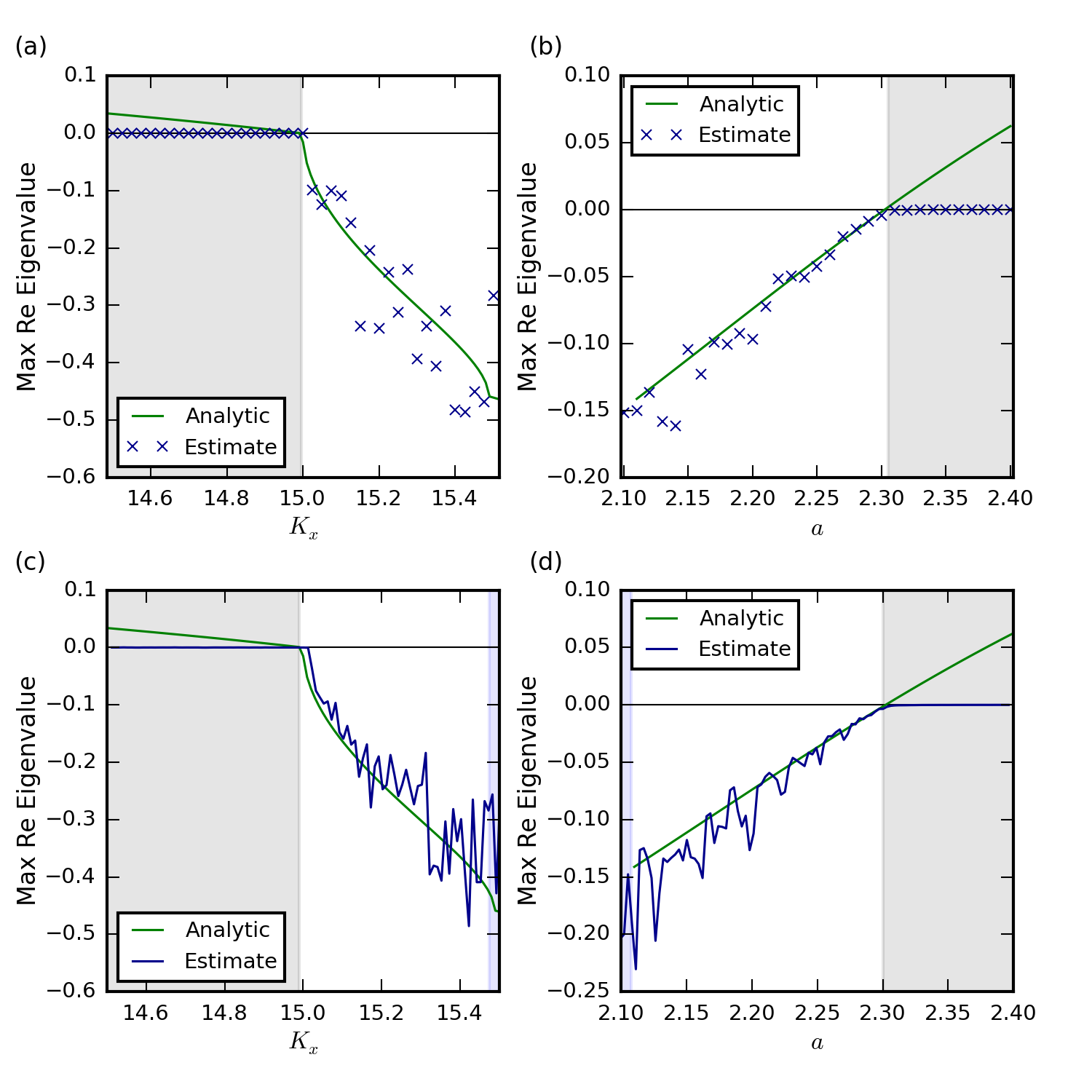}
    \caption{Eigenvalue estimations in the Rosenzweig-MacArthur with quadratic mortality and diffusion on the six-patch network. The steady state under consideration loses stability due to a Fold (panels a,c) and Hopf (b,d) bifurcations. The leading eigenvalue was reconstructed from simulation runs with fixed parameters (a,b) and slowly changing parameters (c,d).    
    The accuracy of the estimates improves as the system approaches the bifurcation. (Estimated values are shown in the middle of the sampling window, indicated in blue. See appendix for details)
    }
    \label{fig:saddle_node}
\end{figure}

So far we studied systems that were designed using the generalized modelling approach. We complement this by the analysis of a well established ecological model, the Rosenzweig-MacArthur predator-prey model \cite{rosenzweig1963graphical} with quadratic mortality and diffusion on the 6-patch network (see appendix for details). The results of Jacobian reconstruction (Fig.~7) show that the leading eigenvalue can be recovered with reasonable accuracy, again the accuracy of the estimate improves significantly as the system approaches the bifurcation point. 

\section{Summary and discussion \label{secDiscussion}}
In this paper we expanded on previous work by Honerkamp, van Kampen, Steuer and others to derive a closed form expression for the reconstruction of Jacobian matrices from time series data. 

For illustration we applied the mathematical formula to the 
an ecological meta-foodweb model. This example illustrated that a relatively robust reconstruction of the leading eigenvalue of the Jacobian is possible even in a strongly nonlinear multi-layer network with dynamics on multiple timescales. However, the example also revealed that the required amount of data is still prohibitive for the ecological application. 

There is reason to believe that future research and particular a deeper mathematical understanding of the Jacobian reconstruction 
can significantly reduce the required amount of data. Particularly interesting in this respect is the observed increase in accuracy close to the bifurcation point. A promising goal for future exploration would be to understand how far this region of heightened accuracy extends. If a real world application under consideration is already close to bifurcation one might find that much less data is required to reach the desired accuracy.

Alternatively, it might be possible to reduce the data demand by optimizing the sampling scheme. One can envision an iterative scheme, similar to \cite{aufderheide2013predict}, where a small number of samples is used to find an initial estimate of the Jacobian. Using the initial estimate one could then identify the relevant timescales and important entries in the covariance matrix and optimize the sampling effort accordingly.

A third alternative may be to use additional information that may be available. The advantage of our jacobian-based approach is that it can take advantage of additional knowledge on the systems that may be available in ecological applications, such as closure exponents or the predator-dependence of the predation rate. The previous example \cite{lade2012early} suggests that the use of this information may reduce the data demand considerably.      

Meanwhile the method proposed here may be useful in fields where data is more readily available, such as studies of metabolism, power grids, or economic data. In the study of metabolism Jacobian reconstruction is already frequently used \cite{khatibipour2018jacly}, for this application the present work yields an analytic closed form solution to a problem that is so far solved by machine learning methods. For power grids, reconstructing Jacobians may be particularly interesting because it could yield deeper insights into the functioning of the system in addition to providing an early warning signal. For economics it may be particularly interesting that the Jacobian can be seen as a representation of causality in the system. Closed form jacobian reconstruction thus offers a way to infer causality from correlation and is guaranteed to be exact in the large data limit.     
An important caveat is that the method requires some additional information to avoid underdeterminedness. However, as illustrated here, it is already sufficient if we can identify $N(N-1)/2$ variable pairs that do not interact directly among the $N$ variables of the system. In networks language, this is equivalent to saying that the mean degree $z$ of the network of interactions must obey 
\begin{equation}
    z<\frac{N-1}{2}\, ,
\end{equation}
a condition that should be easy to satisfy in many applications and becomes easier to meet in larger systems.  

In summary, we find that Jacobian reconstruction is a promising approach to the analysis of complex systems near critical transitions, although the data requirements presently still limit its applicability. We expect that the closed form solution derived in the present paper inspires future mathematical work to alleviate these requirements. 

\section*{Acknowledgements}
The authors thank Caio Lucidius N.~Azevedo and Nicolas Verschueren van Rees for valuable pointers to the algebra literature. This work was supported by Deutsche Forschungsgemeinschaft Research Unit FOR 1742, under grant contract number Dr300/13-2 and by the EPSRC (EP/N034384/1). 


\appendix

\section{Generation of noisy timeseries \label{appendixEMmethod}}
The stochastic system is modeled as an Itō process. The calculation is done by using the Euler-Maruyama method \cite{kloeden1992approximation}
\begin{align}
	x(t+dt) &= x(t) + \dot x(t) \, \text dt + a \sqrt{x(t)} \, \text dW(t) \;,
\end{align}
where $\text dW(t)$ is the increment of a Wiener process with a normal distribution around the mean 0 and the standard deviation $\sqrt{\text dt}$. The amplitude of the noise is proportional to the square root of the population size if we suppose that the noise is due to intrinsic fluctuations of the population dynamics due to stochastic birth and death processes. While realistic for ecological systems, it is important to note that this does not match the stochastic behaviour (additive noise) that was assumed in derivation of the relationship between the co-variance and jacobian matrices. The subsequent results demonstrate that this assumption in the formulation of the method is not a barrier to its application.
For the time series we used the noise strength $a=0.01$ and different step sizes $\text dt$. The number of used time steps and step sizes is listed in table~\ref{tab:solver_pars}.

\section{Jacobian-Covariance relationship \label{appendixCovJ}}
We summarize the derivation of the Jacobian-Covariance relationship following the presentation in \cite{steuer2003observing}.
The response of the system to small fluctuations of the variables around the equilibrium values can be approximated by
\begin{align}
\frac{\mathrm{d}}{\mathrm{d}t} \mathbf{X} = \mathbf{J} \mathbf{X},
\end{align}
We can model the system with noise using a Langevin-type equation
\begin{align}
    \frac{\mathrm{d}X_i}{\mathrm{d}t}=\sum_j J_{ij} X_j +\sqrt{2D_i} \, \xi_i(t),
\end{align}
where $\xi_i(t)$ is Gaussian white noise, with zero mean and unit variance and $D_i$ is the mean amplitude of fluctuations. 

The corresponding stationary Fokker-Planck equation for the state probability distribution $P(\mathbf{X})$ is 
\begin{align}
    -\sum_{ij}J_{ij}\frac{\partial}{\partial X_i} X_j P +\sum_{ij}D_{ij}\frac{\partial^2P}{\partial X_i \partial X_j}=0 \label{eq:fp}.
\end{align}
Multiplying Eq.~\eqref{eq:fp} by $X_k X_l$ and integrating gives 
\begin{align}
\sum_j J_{kj} \left<X_lX_j\right> + \sum_j J_{lj} \left<X_kX_j\right> +2D_{kl}=0,
\label{eq:cvelement}
\end{align}
where $\left<X_iX_j\right>$ is the co-variance of variables $X_i$ and $X_j$. 
Equation~\eqref{eq:cvelement} can be written in matrix form
\begin{align}
    \mathbf{J\Gamma}+\mathbf{\Gamma J^T} = -2\mathbf{D},
\end{align}
where $\mathbf{\Gamma}$ is the co-variance matrix with entries $\Gamma_{ij}=\left<X_iX_j\right>$ \cite{vankampen2007stochastic}.

\section{Vectorization of matrix products \label{appendixVecKronecker}}
Following \cite{magnus2019matrix} we consider a product of three matrices ${\bf M}={\bf XYZ}$. 
We find the vectorization of $\bf M$ by stacking its columns, i.e.
\begin{equation}
{\rm vec}(\bf M)=\left(\begin{array}{c}\boldsymbol{m_1}\\ \boldsymbol{m_2}\\ \vdots \end{array}\right)    
\end{equation}
where $\boldsymbol{m_i}$ is the $i$-th column of $\bf M$. We can obtain the $i$-th column of $M$ by replacing $\bf Z$ by its $i$-th column $\boldsymbol z_i$, which yields 
\begin{equation}
{\bf XY}\boldsymbol{z_i} =  \sum_j {\bf X} \boldsymbol{y_j} Z_{j,i} =  \sum_j Z_{j,i} {\bf X} \boldsymbol{y_j} 
\end{equation}
where $\boldsymbol{y_j}$ is the $j$-th column of $\bf Y$.
The sum on the right-hand-side is also the product of the factors
\begin{equation}
(Z_{1,i}{\bf X},Z_{2,i}{\bf X},\ldots) = \boldsymbol{z_i}^{\rm T} \otimes {\bf X}
\end{equation}
and 
\begin{equation}
\left(\begin{array}{c} \boldsymbol{y_1}\\ \boldsymbol{y_2} \\ \vdots \end{array}\right) = {\rm vec}({\bf Y}) 
\end{equation}
Hence 
\begin{equation}
{\bf XY}\boldsymbol{z_i} = (\boldsymbol{z_i}^{\rm T} \otimes {\bf X}){\rm vec}({\bf Y})      
\end{equation}
Stacking these equations for the different values of $i$ yields
\begin{equation}
\left(\begin{array}{c}
{\bf XY}\boldsymbol{z_1} \\ {\bf XY}\boldsymbol{z_2} \\ \vdots
\end{array}\right)
= 
\left(\begin{array}{c}
(\boldsymbol{z_1}^{\rm T} \otimes {\bf X}){\rm vec}({\bf Y}) \\      
(\boldsymbol{z_2}^{\rm T} \otimes {\bf X}){\rm vec}({\bf Y}) \\
\vdots
\end{array}\right)
\end{equation}
and hence the result
\begin{equation}
{\rm vec}(\bf XYZ) = ({\bf Z}^{\rm T} \otimes {\bf X}){\rm vec}({\bf Y})      
\end{equation}

\section{Pseudoinverse of overdetermined system \label{appendixPseudoinverse}}
Consider a system of the form 
\begin{equation}
{\bf B}\boldsymbol{v}=\boldsymbol{w}
\end{equation}
If ${\bf A}$ has a row-dimension that is greater than the column dimension of $\bf A$ this is an overdetermined system, so for a given $\bf A$ and $\boldsymbol w$, we cannot generally expect that there is a $\boldsymbol v$ that solves the equation. Hence for any $\boldsymbol v$ there will be some residue $\boldsymbol x$:
\begin{equation}
{\bf B}\boldsymbol{v}-\boldsymbol{w}=\boldsymbol{x}
\end{equation}
Our aim is now to find the $\boldsymbol v$ such that the residue $\boldsymbol x$ is minimized. Specifically we seek to minimize the euclidean norm 
\begin{equation}
|\boldsymbol x|^2 = \boldsymbol x^{\rm T} \boldsymbol x
\end{equation}
This expression has a unique minimum at which the gradient vanishes \cite{trefethen1997numerical}. Hence we can find the desired $\boldsymbol v$ by demanding 
\begin{equation}
0  = \nabla  (\boldsymbol x^{\rm T} \boldsymbol x) 
\end{equation}
where $\nabla= (\partial/\partial v_1, \partial/\partial v_2, \ldots )^{\rm T}$. 
Transforming this equation we find 
\begin{eqnarray}
0  &=& 2 (\nabla  \boldsymbol x^{\rm T}) \boldsymbol x \\
   &=& \left(\nabla  (\boldsymbol{v}^{\rm T}{\bf B}^{\rm T}-\boldsymbol{w}^{\rm T})\right) \boldsymbol x  \\
   &=& {\bf B}^{\rm T} \boldsymbol x \\
   &=& {\bf B}^{\rm T}{\bf B}\boldsymbol{v}-{\bf B}^{\rm T}\boldsymbol{w}
\end{eqnarray}
We can write this condition as 
\begin{equation}
{\bf B}^{\rm T}{\bf B}\boldsymbol{v}={\bf B}^{\rm T}\boldsymbol{w}    
\end{equation}
on the left-hand-side the square matrix ${\bf B}^{\rm T}{\bf B}$ appears. In contrast to the rectangular matrix $\bf B$ this matrix can typically inverted. 
Hence we can multiply the inverse from the left to obtain the desired formula
\begin{equation}
\boldsymbol{v}=\left({\bf B}^{\rm T}{\bf B}\right)^{-1}{\bf B}^{\rm T}\boldsymbol{w}    
\end{equation}

\section{Metafoodweb model \label{appendixFWmodel}}
Our aim in this section is to formulate a model that can serve as a test case 
for the Jacobian reconstruction method. For this purpose we want a large, complex, and nonlinear dynamical network model where the Jacobian matrix is nevertheless analytically accessible,  

The model consists of a spatial network of $N$ habitat patches linked by avenues of species dispersal. Each patch harbours a complex food web, consisting of $P$ nodes that represent populations of different species, which are linked by predator-prey interactions. 

The variables of the model $X_{si}$ denotes the population density of species $s$ on patch $i$. For simplicity we assume that all patches are identical. The population dynamics is given by the general form 
\begin{equation}
\begin{array}{r c l l}
\dot{X}_{si} &=& G(X_{si})+\epsilon F(X_{si},T_{si}) &                             \mbox{Local gains}   \\
            & &  -\left(\sum_p \frac{c_{sp}X_{si}}{T_{pi}}F(X_{pi},T_{pi})\right) - M(X_{si}) & \mbox{Local losses}\\ 
              & & -d_sk_iX_{si}+d_s\sum_j A_{ij} X_{sj} &   {\mbox{Dispersal}}\end{array}
\label{eqgeneral}
\end{equation}
Let's unpack this equation. The first of the gain terms, $G$ represents primary production of biomass growth, e.g.~by photosynthesis. We assume that this term is zero for all predator species. By contrast, the second term with the functional response $F$, which describes growth by predation, is assumed to be zero for primary producers. 
This second term represents growth by predation, which depends on the density of the predator, and the total density of prey, $T_{si}$ that fits the predators diet. The total prey density for predator $s$ in patch $i$ can be written as 
\begin{equation}
T_{si} = \sum_p c_{ps} X_{pi}  
\end{equation}
where $c_{ps}$ is the relative contribution that $p$ makes to the diet of $s$. For example $c_{ps}=1$ if species $p$ is easy prey for species $s$, but $c_{ps}=0$ if $s$ cannot prey on $p$. 
The first loss term in Eq.~(\ref{eqgeneral}) captures losses by predation, where we assumed that the biomass uptake by a predator is assigned to the predator's prey species according to their diet. The function $M(X_{si})$ describes losses due to natural (i.e.~non-predatory) mortality.     

The final two terms describe the effect of diffusive emigration from and immigration to the patch, with a species dependent diffusion rate $d_s$. The matrix $\bf A$ is the adjacency matrix of the geographic network, and $k_i$ the degree of patch $i$. With this diffusive coupling there is always a solution where all patches in the networks have the same densities in all populations
\begin{equation}
{X_{si}}^* = {X_s}^*
\end{equation}
In the following we call this steady state the homogeneous state. 

We note that at this stage the functions $G$, $F$ and $M$ are still unspecified. Using the approach of Brechtel  et al.~\cite{brechtel2018master} the Jacobian matrix for this type of model can be computed analytically. The result is a Jacobian $\bf J$ that still contains a set unknown, but ecologically interpretable parameters, that describe properties of the unspecified functions in the model. In \cite{brechtel2018master} we showed that the Jacobian in the homogeneous state can be written as 
\begin{equation}
{\bf J} = {\bf I} \otimes {\bf P} - {\bf L} \otimes {\bf K}    
\end{equation}
Here $\bf I$ is an $N \times N$ identity matrix, $\bf P$ is the Jacobian of a single isolated patch, $\bf L$ is the Laplacian matrix of the geographical network and $\bf K$ is a diagonal coupling matrix
\begin{equation}
{\bf K} = \left(\begin{array}{c c c}
d_1 \\
& d_2 \\
& & \ddots 
\end{array} \right)    
\end{equation}

Following the procedure in \cite{brechtel2018master} one can show that the eigenvalues $\lambda$ of $\bf J$ can be computed as 
\begin{equation}
\lambda = {\rm Ev}\left({\bf P}-\kappa {\bf K}\right)    
\end{equation}
where $\kappa$ is a laplacian eigenvalue of the geographic network. Solving this equation for every $\kappa$ yields the complete set of eigenvalues of $\bf J$. 

The patch Jacobian $\bf P$ for this model was already derived in \cite{gross2006generalized} in a non-spatial context. In summary these results allow for the very convenient computation of a broad class of fairly realistic meta-foodweb models. Once we have found the desired bifurcations in a food web it is straightforward to specify the unspecified function in the model to lie at a specific point of the generalized model bifurcation diagram. In this way meta-foodwebs that exhibit the desired transitions can be generated very efficiently. 

For most of the analysis we use the functions
\begin{eqnarray}
 G(X_s) &=& \alpha_s  {X_s}^{\phi_s} \\
 F(X_s,T_s) &=& \alpha_s  \frac{(1+K_s){X_s}^{\psi_s}T_s }{T_s + K_s} \\
 M(X_s) &=& \alpha_s {X_s}^{\mu_s}
\end{eqnarray}
here $\alpha_s$ is the species biomass turnover rate, which we calculate from its trophic position using an allometric scaling law; $\phi_s$ is the elasticity of primary production, $\mu_s$ is the elasticity of mortality, $\psi_s$ is the elasticity of predtion with respect to predator density, and $K_s$ is a halfsaturation constant which we set to $K_s=\gamma_s/(1-\gamma_s)$ where $\gamma_s$ is the desired elasticity of predation with respect to the prey. 

For the simulations with the Rosenzweig-MacArthur model we 
used 
\begin{eqnarray}
\dot{X}_{1i} &=& rX_{1i} \left(1 - \frac{X_{1i}}{c} \right) 
- \frac{a X_{1i} X_{2i}}{1+ahX_{1i}} + \sum_jd_1(X_{1j}-X_{1i}) \\
\dot{X}_{2i} &=& \epsilon \frac{a X_{1i} X_{2i}}{1+ahX_{1i}} -mX_{2i} - (wX_{2i})^2 + \sum_jd_2(X_{1j}-X_{1i}) 
\end{eqnarray}

The full set of parameters used in all simulations is shown in the  tables below. 
\begin{table}[htb]
\begin{center}
\begin{tabular}{|c||c|c|c|c|c|c|c|c|}
    \hline
    Figure & 3 & 5 & 7 (a), (b) & 7 (c), (d) & 6 (a), (b) & 6 (c), (d) \\
	\hline
	\hline
	$T_\text{max}$ & 200 & 4 200 & 2 000 & 80 000 & 20 000 & 50 000\\
	\hline
	steps & $2\cdot10^5$ & $4.2 \cdot 10^6$ & $2\cdot 10^5$ & $2\cdot 10^6$ & $2\cdot 10^6$ & $5\cdot10^{7}(*)$\\
	\hline
	$\text dt$ & $10^{-3}$ & $10^{-3}$ & $10^{-2}$ & $10^{-2}$ & $10^{-2}$ & $10^{-3}$\\
	\hline
\end{tabular}
\end{center}
\caption{The parameters of the solver. The Length of the simulation $T_\text{max}$ and the number of timesteps.
$(*)$ In the case of figure~\ref{fig:bigger_system} (c), (d) only $2\cdot10^{6}$, that means every 25th step, of the time steps were used for the eigenvalues estimation.}
\label{tab:solver_pars}
\end{table}

\begin{table*}[htb]
\begin{tabular}{l|p{0.75\textwidth}}
\hline
Parameter & Interpretation\\
\hline
\hline
Exponent& \\
\hline
$\phi$ & Sensitivity of primary production to own population density\\
$\gamma$ & Sensitivity of predation to total available prey density \\   	
$\psi$ & Sensitivity of predation to predator density\\
$\mu$ & Exponent of closure\\
\hline
\hline
Scale & \\
\hline
${\alpha}$ & Biomass flow\\
$\sigma$ & Fraction of biomass loss due to predation\\
$\tilde{\sigma}$ & Fraction of biomass loss due to respiration\\
$\beta$ & Relative contribution to biomass loss due to predation by a certain predator\\
$\delta$ & Fraction of local growth by predation\\  
$\tilde{\delta}$ & Fraction of local growth by primary production\\  
$\chi$ & Relative contribution of population as prey to a certain predator\\
\hline
\end{tabular}
\caption{Generalized parameters used to describe the meta-foodweb.}
\label{tab:gen_parameters}
\end{table*}

\begin{table}[htb]
    \centering
    \begin{tabular}{|c||c|c|c|c|c|c|c|c|}
	    \hline
	    $i$ & $\alpha_i$ & $\phi_i$ & $\mu_i$ & $\psi_i$ & $\gamma_i$ & $\sigma_i$ & $\delta_i$ & $d_i$\\
	    \hline
	    \hline
	    1 & 10 & $\phi$ & 2 & - & - & 0.9 & 0 & 3 \\
	    \hline
	    2 & 3 & - & 2 & 1 & $\gamma$ & 0 & 1 & 10 \\
	    \hline
    \end{tabular}
    \caption{Parameters used for the two species predator-prey system.}
    \label{tab:2S_parameters}
\end{table}

\begin{table}[htb]
    \centering
    \begin{tabular}{|c|c|c|}
        \hline
        $(\phi, \gamma)_\text{start}$ & $(\phi, \gamma)_\text{end}$ & Bifurcation\\
        \hline
        \hline
        (0.25, 0.07) & (0.30, 0.02) & Transcritical\\
        \hline
        (0.30, 0.10) & (0.35, 0.05) & Transcritical \\
        \hline
        (0.35, 0.13) & (0.40, 0.08) & Transcritical \\
        \hline
        (0.40, 0.16) & (0.45, 0.11) & Transcritical \\
        \hline
        (0.45, 0.18) & (0.50, 0.13) & Transcritical \\
        \hline
        (0.50, 0.21) & (0.55, 0.16) & Transcritical \\
        \hline
        (0.55, 0.24) & (0.60, 0.19) & Turing \\
        \hline
        (0.60, 0.27) & (0.65, 0.22) & Turing \\
        \hline
        (0.65, 0.30) & (0.70, 0.25) & Turing \\
        \hline
        (0.70, 0.35) & (0.75, 0.30) & Turing \\
        \hline
        (0.75, 0.38) & (0.80, 0.33) & Turing \\
        \hline
        (0.80, 0.41) & (0.85, 0.36) & Turing \\
        \hline
        (0.85, 0.46) & (0.90, 0.41) & Turing \\
        \hline
        (0.90, 0.51) & (0.95, 0.46) & Hopf \\
        \hline
        (0.95, 0.56) & (1.00, 0.51) & Hopf \\
        \hline
    \end{tabular}
    \caption{Start and end points of the parameter trajectories used in the error statistics plot (Fig.~4) together with the corresponding bifurcation.}
    \label{tab:2S_parameters2}
\end{table}

\begin{table}[htb]
\begin{center}
\begin{tabular}{|c|c|c|c|c|c|c|c|c|}
	\hline
	$r$ & $c$ & $\epsilon$ & $m$ & $w$ & $a$ & $h$ & $d_1$ & $d_2$\\
	\hline
	2 & 10 & 1 & 1 & 0.2 & 3 & 0.5 & 1 & 1 \\
	\hline
\end{tabular}
\end{center}
\caption{The parameter set used in the modified Rosenzweig-MacArthur model to find a saddle-node bifurcation.}
\label{tab:SN_parameters}
\end{table}

\begin{table}[htb]
    \centering
    \begin{tabular}{|c||c|c|c|c|c|c|c|c|}
	    \hline
	    $i$ & $\alpha_i$ & $\phi_i$ & $\mu_i$ & $\psi_i$ & $\gamma_i$ & $\sigma_i$ & $\delta_i$ & $d_i$\\
	    \hline
	    \hline
	    1 & 0.10 & - & 1 & 1 & $\gamma$ & 0 & 1 & 0.01 \\
	    \hline
	    2 & 0.45 & - & 1 & 1 & $\gamma$ & 0.6 & 1 & 0.01 \\
	    \hline
	    3 & 0.12 & - & 1 & 1 & $\gamma$ & 0 & 1 & 0.01\\
	    \hline
	    4 & 0.75 & $\phi$ & 1 & - & - & 0.6 & 0 & 0.01 \\
	    \hline
    \end{tabular}\\[2mm]
    \begin{tabular}{|c||c|c|c|c|}
	    \hline
	    $i,j$ & $1,2$ & $1,4$ & $2,4$ & $3,2$ \\
	    \hline
	    \hline
	    $\chi_{ij}$ & $0.5$ & $0.5$ & $1$ & $1$ \\
	    \hline
	    $\beta_{ij}$ & $1/3$ & $1/3$ & $2/3$ & $2/3$ \\
	    \hline
    \end{tabular}
    \caption{Parameters used for the 4 species food web system. $\phi_4$ is used as the bifurcation parameter, all other parameters are kept constant. For the the relative gain $\chi_{ij}$ and loss $\beta_{ij}$ only the nonzero entries are shown. In the used example $A_{ij} = \chi_{ij}$. If $\phi$ is used as the bifurcation parameter we choose $\gamma=0.95$. If instead $\gamma$ is used as the bifurcation parameter we choose $\phi=0.93$.}
    \label{tab:4S_parameters}
\end{table}

\end{document}